\documentclass{agujournal2019}
\usepackage{url} 

\usepackage{amssymb}

\def\arraystretch{1.5}

\journalname{JGR: Solid Earth}

\begin{document}

%
%

\title{Topological properties of epidemic aftershock processes}

%
%




\authors{Jordi Bar\'o\affil{1}}


\affiliation{1}{Centre for Mathematical Research,
Campus de Bellaterra, Edifici C	 
08193 Bellaterra, Barcelona, Spain}




\correspondingauthor{Jordi Bar\'o}{jbaro@crm.cat}




\begin{keypoints}
\item Epidemic aftershock models are interpreted as branching processes with causal-links.
\item All topological properties of a simple epidemic model are determined by only two parameters.
\item If valid, model predictions can be used to classify natural and artificial catalogs.  
\end{keypoints}

%
%


\begin{abstract}
Earthquakes in seismological catalogs and acoustic emission events in lab experiments can be statistically described as a linear Hawkes point process, where the spatio-temporal rate of events is a linear superposition of background intensity and the aftershock clusters triggered by preceding activity.
Traditionally, statistical seismology has interpreted this model as the outcome of an epidemic branching process, where one-to-one causal links can be established between mainshocks and aftershocks. 
Declustering techniques have been used to infer the underlying triggering trees and relate their topological properties with epidemic branching models. Here, we review how the standard Epidemic Type Aftershock Sequence (ETAS) model extends from the Galton-Watson (GW) branching processes and bridges two extreme cases: Poisson sampling and scale-free power-law trees. We report the most essential topological properties expected in GW epidemic trees: the branching probability, the distribution of tree size, the expected family size, and the relation between average leaf-depth and tree size.
We find that such topological properties depend exclusively on two sampling parameters of the standard ETAS model: the average branching ratio $N_b$ and the exponent ratio $\alpha/b$ determining the branching probability distribution. From these results, one can use the memory-less GW as a null-model for empirical triggering processes and assess the validity of the ETAS model to reproduce the statistics of natural and artificial catalogs. 
\end{abstract}

%
%

 \section{Introduction}

The concept of aftershocks is traditionally associated with seismology \cite{Utsu1995}, but similar phenomena have been observed in other natural systems and are common in many mechanical processes in rocks, composites and porous materials \cite{Benioff1951,Hirata1987,Baro2013,Ribeiro2015,
Davidsen2017}.
Aftershocks are identified in sequences of point events as a sudden increase of the activity causally linked to a previous event ---usually stronger--- called a mainshock.
The empirical Omori-Utsu law \cite{Utsu1995} describes the common temporal evolution of the number of aftershocks after time $\tau$ since a mainshock of magnitude $m_{\mathrm{MS}}$ as:
  \begin{linenomath*}
  \begin{equation}
    N_{AS}(\tau| m_{\mathrm{MS}}) \sim \frac{k 10^{m_{\mathrm{MS}}}}{(C+\tau)^p},
    \label{eq:Omori}
  \end{equation}
  \end{linenomath*}
  where $p$ is a power-law exponent usually close to 1 and $k$ a productivity factor.
Additionally, aftershocks are spatially clustered, usually according to power-law decay with distance to the mainshock $r := ||\mathbf{r}-\mathbf{r_{\mathrm{MS}}}||$, being $\mathbf{r}$ and $\mathbf{r_{\mathrm{MS}}}$ the locations of the aftershock and the mainshock respectively \cite{Guo1995}. Different sophistications have been proposed accounting for more precise observations such as anisotropic spatial drift of the aftershock production and non-factorizable magnitude dependencies \cite{Ogata2006},  generalized scaling forms \cite{Vere2005,Saichev2005,Davidsen2016} or more complex temporal decay forms \cite{Davidsen2017,Baro2017}. Such details are excluded from the following mathematical and numerical developments but will be recovered in the discussion of the results.\\
\smallskip
\begin{figure}
\includegraphics[scale=1]{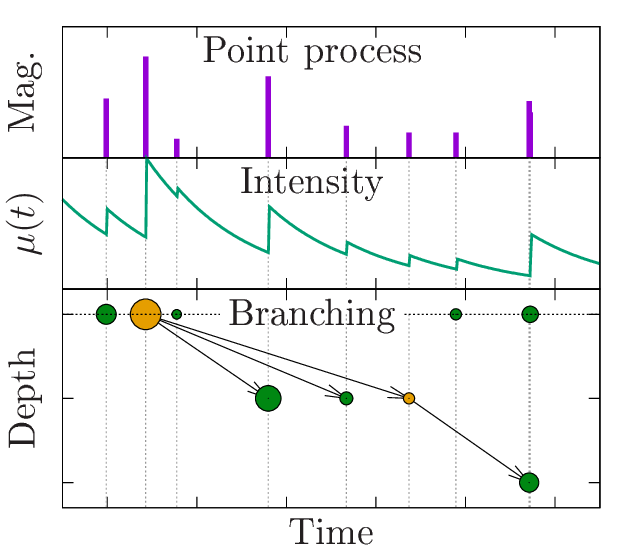}
\caption{\label{fig:pp}
Schematic representation of a temporal sequence of aftershocks as events in a marked point process (top); the intensity can be inferred under the assumption of a linear Hawkes process (center); and interpreted as a branching process (bottom). Background events occur at Depth = 0. Dark green circles represent leaves or singlets if Depth = 0. }
\end{figure}
Due to the complexity of the seismogenic mechanisms, statistical seismology considers all earthquakes ---mainshocks and aftershocks--- as non-isolated stochastic events in space and time, rather than the outcome of deterministic mechanical processes. 
Mainshocks and aftershock sequences from different mainshocks coexist  in the same regions and temporal windows. 
As consequence, all events are customarily interpreted as point events in a stochastic point process, determined by an intensity $\mu(t, \mathbf{r}, m)$ accounting for the instantaneous probability of finding an event defined by a mark ---in this case, the earthquake magnitude $m$--- at a spatiotemporal ($t, \mathbf{r}$) location. 
Simple proportional hazard models consider events to be independent, with a space-time dependent $\mu$ \cite{Varotsos1996}. More sophisticated hazard models take into account the correlations between events caused by aftershock production \cite{Vere1982, Ogata1988, Michael1997, Zhuang2002, Turcotte2007}. The most simple aftershock model is to consider a linear Hawkes self-exciting process~\cite{Hawkes1974} incorporating the observations of statistical seismology. In a linear Hawkes process, the intensity can be expressed as the linear superposition of a background rate $ \mu_0$ and the individual contribution of all previous events $\lbrace i\rbrace$ with a triggering term $\Psi_i$:
  \begin{linenomath*}
  \begin{equation}
    \mu(t, \mathbf{r}, m) = \mu_0(t, \mathbf{r}, m) + \sum_{i|t_i < t} \Psi_i( m,t,\mathbf{r} |m_i, t_i,\mathbf{r_i}).
    \label{eq:Hawkes}
  \end{equation}
  \end{linenomath*}
Notice that the linear Hawkes model assumes an additive contribution for each event in the intensity, meaning that all events can generate aftershocks. This effective stochastic process is often interpreted as the result of an epidemic or branching process \cite{Vere1966,Ogata1988,Saichev2004,Turcotte2007}. Fig.~\ref{fig:pp} shows a schematic representation of a sequence of point events interpreted as the outcome of a history-dependent intensity, or Hawkes process, and its representation as a branching process. The branching process is strictly constituted by two distinct categories of events: background independent events generated entirely by the background rate ($\mu_0$ in Eq.~(\ref{eq:Hawkes})), and triggered events, caused by a unique preceding parent event $i$, through its individual and independent contribution to the intensity ($\Psi_i$ in Eq.~(\ref{eq:Hawkes})), represented as arrows in Fig.~\ref{fig:pp}. Under the branching process assumption, earthquake catalogs are built as \textit{branching forests}, linear superpositions of independent topological objects that here we call \textit{triggering trees}. Each triggering tree is a structure of causally connected events initiated by a background event, the root of the tree (at Depth = 0 in Fig.~\ref{fig:pp}), that can trigger a number $N_1$ of events in a first generation of aftershocks (Depth = 1 in Fig.~\ref{fig:pp}). In its turn, each of the $N_{d}$ events in the $d$-th generation of aftershocks can trigger events in the $(d+1)$-th generation of aftershocks, and so on. We call \textit{leaves} those events with no offspring, extinguishing a branch (dark-green circles in Fig.~\ref{fig:pp}). \textit{Singlets} are background events which are also leaves, i.e. have no offspring. The triggering tree is extinguished when all events in a maximum generational depth $D$ are leaves.\\
\medskip
%
Assuming the validity of the branching process approach, the identification of triggering trees would provide valuable topological information of the branching and enable the direct measurement of the triggering kernel $\Psi_i$ \cite{Zhuang2004, Davidsen2017}. Hence, this representation of aftershock sequences is a useful approach, whether we can argue if strictly valid. 
 Whilst the stochastic point process resulting from the branching process can be represented as a linear Hawkes process, the linear Hawkes processes do not require the existence of explicit one-to-one causal links between events. 
In theory, univocal causal links in mechanical interactive systems can be defined from an energy stability point of view. This can be shown in micromechanical models such as the viscoelastic democratic fiber bundle model~\cite{Baro2018a}.
In field studies, however,  we have a limited capability to identify such a deterministic process. 
The bare statistical analysis of earthquake catalogs do not determine the explicit causal links but provide, instead, an assessment of its point process representation (\ref{eq:Hawkes}) where all terms contribute to $\mu$ with their specific weight. 
Advanced declustering techniques, either stochastic \cite{Zhuang2004} or based on the nearest neighbor distance \cite{Baiesi2004, Zaliapin2013a, Zaliapin2013b} can be used to infer the most plausible one-to-one causal structure.
This classification is never entirely free of uncertainty
 \cite{Zhuang2002} but appears to be reliable when tested against synthetic catalogs~\cite{Zaliapin2013a}.  
Notice that the topological concepts discussed in the following sections are only valid under the branching process assumption and have no correspondence to the more general point process description.\\
\medskip
Recently, the analysis of seismological catalogs as branching processes revealed significant deviations between the reconstructed clusters or triggering trees, and the branching model expectations in southern California~\cite{Zaliapin2013b}. The same authors suggested a regional classification based upon such inconsistencies in the topological properties. Thereupon, aftershock sequences or clusters were classified into two distinct categories: bursts and swarms. \textit{Burst-like} clusters were defined as clusters of events with shallow generational depth ($d$), where most activity is accumulated in the first generation of aftershocks, mostly a consequence of significantly strong mainshocks. The burst-like activity was found to be compatible with the numerical results of the Epidemic Aftershock Sequence (ETAS) model, defined in the next section, by imposing the parameters fitted from field catalogs. Burst-like sequences were linked to tectonic settings with low heat flow. On the contrary,  \textit{swarm-like} clusters designate aftershock sequences with deeper generational depths, usually growing with the size of the swarm. The swarm-like activity was not predicted by the ETAS model with the fitted parameters and was found to match those regions with high heat flow.
The results by~\citeA{Zaliapin2013b} paved the road for a new analytical methodology based upon the topological statistics of triggering trees. 
This methodology will potentially lead to a better understanding of the seismogenic mechanisms behind triggering processes and improve the accuracy of stochastic point-process models with potential applicability to hazard assessment~\cite{Field2014}. Overall, the study of topological properties of triggering trees will foreseeably be gaining more popularity in the following years, thanks to the improved refinement of event detection techniques \cite{Shelly2016,Ross2017}. However, few works \cite{Saichev2004,Saichev2005} have addressed the topological properties expected from these hypothetical branching models, and a global picture of the model predictions remains incomplete.\\
\medskip
Here, we will introduce some of the most common topological features used to characterize natural triggering  for the standard ETAS model, defined as a particular Galton-Watson branching model. We will revisit the results by~\citeA{Saichev2005} and introduce general predictions regarding the concepts of family size and average leaf-depths.
These results can be useful as a benchmark to validate ETAS as a null-hypothesis to natural and synthetic catalogs. As a main result, we will prove that the relation between average leaf-depths and cluster sizes depends exclusively on the probabilistic nature of the individual branching ratios, determined in the ETAS model by the ratio of parameters $b/\alpha$ and the average branching rate. This has a direct implication on the separation between swarm-like and burst-like clusters.

\section{Epidemic aftershocks as Galton-Watson branching models}

Epidemic Type Aftershock Sequence (ETAS) models are based on field and lab observations such as the empirical Omori-Utsu law (\ref{eq:Omori}) and the spatial distribution of aftershocks. Here we discuss the standard ETAS model~ \cite{Ogata1988}, simulating a Hawkes process where the
triggering kernel is factorized in its dependencies as:
  \begin{linenomath*}
  \begin{equation}
\Psi_i( m,t,\mathbf{r} |m_i, t_i,\mathbf{r_i}) =\rho_m(m) n_b(m_i)\Psi_{t}(t-t_i)\Psi_{r}(||\mathbf{r}-\mathbf{r_i}||).
\label{eq:ETAS}
  \end{equation}
  \end{linenomath*}
Both the temporal and spatial kernel are normalized to $\int_0^{\infty}  \Psi_{t}(t-t_i) dt = 1$ and $ \int_{\mathbb{R}} \Psi_{r}(||\mathbf{r}-\mathbf{r_i}||) d\mathbf{r}  = 1$ and, therefore, have no effect on the topology of the triggering trees. 
The magnitudes ($m$) of the events are independent and identically distributed (i.i.d.) following the Gutenberg-Richter law \cite{Gutenberg1944}:
  \begin{linenomath*}
  \begin{equation}
\rho_m(m):=10^{-b(m-m_c)}/(b\log(10)),
\label{eq:GR}
  \end{equation}
  \end{linenomath*}
being the magnitude of completeness $m_c$ an effective lower-bound to the distribution. For the sake of simplicity, we consider $m_c$ to be also the minimum magnitude able to generate aftershocks.  We also consider that the number $N_i$ of aftershocks generated by event $i$ is a Poisson number with a characteristic \textit{branching ratio} $n_b:=\mathbb{E}( N_i )$:
  \begin{linenomath*}
  \begin{equation}
\mathbb{P}(N_i=k | n_b) = \frac{n_b^{k}e^{-n_b}}{k!}.
\label{eq:Poisson}
  \end{equation}
  \end{linenomath*}
The branching ratio is given by the aftershock production in Eq.~(\ref{eq:ETAS}): 
\begin{equation}
n_b(m_i):=k_c 10^{\alpha(m_i-m_c)},
\label{eq:productivity}
\end{equation}
reproducing the mainshock-magnitude ($m_i$) dependence in the Omori-Utsu law (\ref{eq:Omori}) with a productivity exponent $\alpha$ usually found between 0.5 and 1 in field~\cite{Utsu1995} and experiments \cite{Baro2013,Davidsen2017}. The term $k_c$ normalizes the aftershock production for $m_i=m_c$. Since magnitudes are independent, the number of aftershocks is equivalent to a random sampling of i.i.d. $n_b$ for all $m_i$ values.
Given Eqs.~(\ref{eq:GR},\ref{eq:productivity}):
\begin{linenomath*}
\begin{equation}
    \rho_{n_b}(n_b)= \frac{b}{\alpha}\frac{1}{k_c} \left({\frac{n_b}{k_c}}\right)^{-\left({\frac{b}{\alpha} + 1}\right)}
   \label{eq:distroNb}
\end{equation}
\end{linenomath*}
where the ratio $\alpha/b$ is bound inside the range $(0,1]$.
From now on we change the notation from $k_c$ to the more convenient average branching ratio $N_b := \langle n_b \rangle= 
 k_c \frac{b}{b-\alpha}$. The number of first generation aftershocks for all events and all $n_b$ is i.i.d. as:
\begin{linenomath*}
\begin{equation}
   \mathbb{P}(N_{i} = k ) = 
\int  \rho_{n_b}(n)\mathbb{P}(N_i=k | n)  dn =
   \frac{b}{\alpha}
   \frac
   {\left({N_b\left({1-\frac{\alpha}{b}}\right)}\right)^{\frac{b}{\alpha}}}
   {k!}
   \Gamma\left({k-\frac{b}{\alpha}, 
   N_b\left({1-\frac{\alpha}{b}}\right)}\right).
   \label{eq:distroN}
\end{equation}
\end{linenomath*}

Therefore, the standard ETAS model is a particular case of a \textit{Galton-Watson} (GW) \textit{process}~\cite{Pitman2006}, where all individual events $i$, background and triggered, have the same probability to trigger a number $N_i$ of events defined by an offspring distribution $\mathbb{P}(N_i=k):=p_1(k)$. In this case, for large $k$, this distribution can be approximated to a power-law  $p_1(k)\sim k^{-\gamma_1}$ with the exponent value $\gamma_1 = {b/\alpha+1}$ inherited from the $n_b$ distribution (\ref{eq:distroNb}). 
Considering $0 < \alpha \leq b$, this exponent values are constraint to $2\leq \gamma_1 $. This result agrees with the reconstructed trees in seismology~\cite{Baiesi2004, Zaliapin2013a}, where $b\approx \alpha$ and the distribution of the number of first generation aftershocks ---called degree distribution by~\citeA{Baiesi2004}--- is a power-law with an exponent $\gamma_1 \approx 2$.\\
\medskip
A singular case of the ETAS model is found for $\alpha = 0$. In that case, the distribution (\ref{eq:distroN}) becomes a Dirac delta around $k_c$: $\lim_{r\to0}\rho(n_b)= \delta(n_b-k_c)$, i.e. $n_b$($= k_c=N_b$) is unique and all events have the same probability of generating aftershocks given by Eq.~(\ref{eq:Poisson}). The ETAS model for $\alpha=0$ is, thus, equivalent to a Poisson Galton-Watson (P-GW) branching process~\cite{Pitman2006}.\\
\medskip

The numerical results in the following sections are obtained through Monte-Carlo generation of a number of aftershock sequences up to $N=10^7$. We use an arbitrary threshold $m_c=1$. The background rate and the spatio-temporal kernel parameters are ignored since they play no role in this study. 


\section{Tree and family sizes}

A fundamental concept for the topological characterization of the branching process is the \textit{tree-size} ($N_T$)  defined here as the total number of members in an extinguished tree for $t\to \infty$: $N_T := \sum_{d=0}^{D} N_d$.  
Notice that, for $N_b>1$, there is a non-zero probability of finding eternal trees ($D \to \infty$) with a non-defined $N_T$. We only consider values $N_b < 1$ through all the discussion, imposing that all trees are extinguished at a finite time.
Again, since all events $i$ are i.i.d.,  $N_T$ is i.i.d. as well, with a given $\mathbb{P}(N_T=K):=p_{T}(K)$. 
In the case of the P-GW, obtained by imposing $\alpha=0$, this distribution is known to be a Borel distribution \cite{Pitman2006}:
  \begin{linenomath*}
  \begin{equation}
p_{\:T}(K| n_b) = \frac{(n_b K)^{K-1}e^{-n_b k}}{K!}.
\label{eq:Borel}
  \end{equation}
  \end{linenomath*}
As a matter of fact, notice that, for large $n_b$ and $K$, this distribution can be approximated by the exponentially tapered power-law: $p_{\:T,n_b}(K) \sim K^{-3/2} e^{-K (1-n_b)^2/2}$. This exponent $\gamma_N= 3/2$ is common in the avalanche size distribution of mean-field models with avalanche dynamics~\cite{Sethna1992,Zapperi1995,Fisher1996,Vespignani1998,Baro2018a}, sometimes regarded as loop-less triggering processes, i.e. P-GW processes.\\ 
\medskip
\begin{figure}
\includegraphics[width=\textwidth]{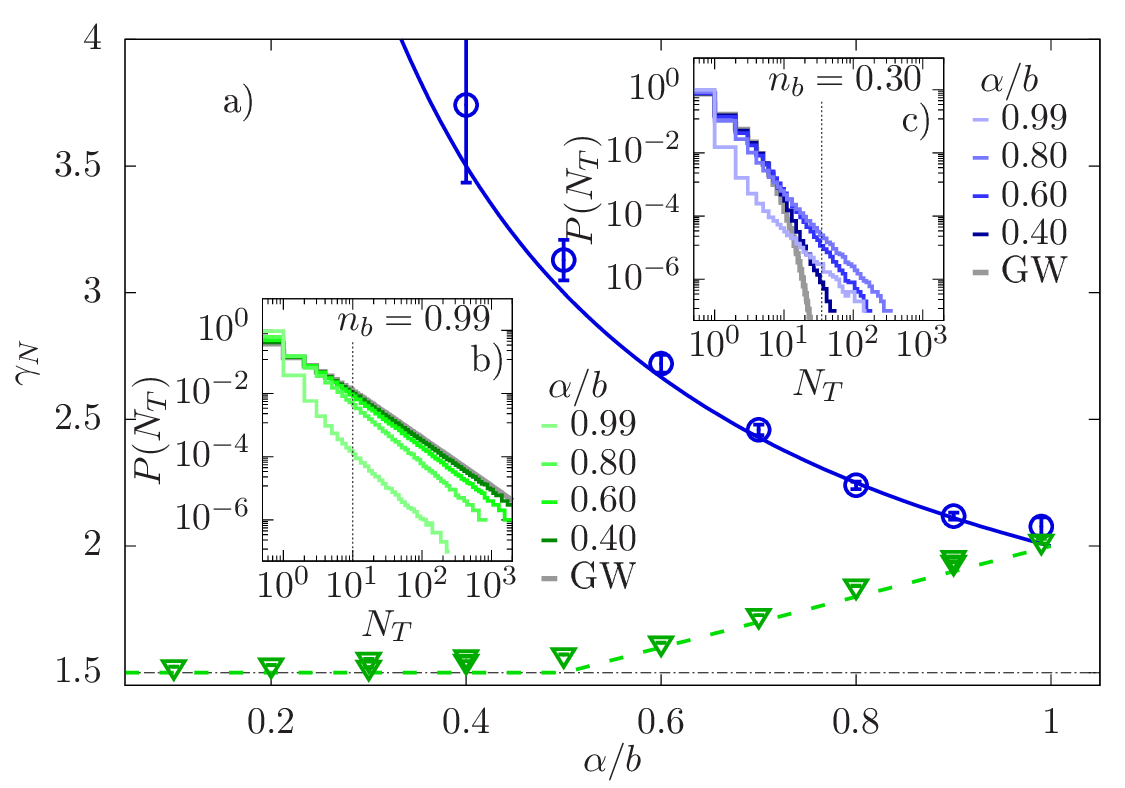}
\caption{\label{fig:pNt} (color on-line) (a) Estimated power-law exponents $\hat{\gamma}_N$ for the distribution of tree sizes ($N_T$) assuming $\mathbb{P}(N_T=k | k >N_{\min}) \sim k^{-\gamma_N}$ for $N_b=0.99$ (green triangles) and  $N_b=0.30$ (blue circles) and different values of $\alpha/b$. Error-bars show an estimate of $\sigma$ of the likelihood function \cite{Baro2012}. The blue and green lines represent the results by~\cite{Saichev2005}.
The dashed black line marks the value $\gamma_N=3/2$ expected in a P-GW process. 
(b,c) Distribution of $N_T$ for selected $\alpha/b$ values and (b) $N_b=0.995$ and (c) $N_b=0.30$ represented in integer exponential binning.  
The exponents are estimated by maximum-likelihood within the interval $N_{\min}<N_{T}<\infty$ taking $N_{\min}=10$ for $N_b = 0.995$ and $N_{\min}=35$ for $N_b = 0.30$ (dashed vertical lines). Grey lines represent the result for a P-GW process with the same $N_b$ (\ref{eq:Borel}).
}
\end{figure}
A decade ago, \citeA{Saichev2005} found the mathematical expression for the distribution of tree sizes ($N_T$). Fig.~\ref{fig:pNt} shows a numerical verification of the results.   The insets (Figs.~\ref{fig:pNt}.b,c) show in log-log scale the distributions of $N_T$ given $N_b = 0.995$ and $N_b = 0.30$ respectively for a selected collection of $\alpha/b$ values, compared with the theoretical result for the P-GW.  Fig.~\ref{fig:pNt}.a shows the power-law exponent ($\hat{\gamma}_N$) estimated by maximum likelihood \cite{Baro2012} for $N_b = 0.99$ and $N_b = 0.30$ as function of $\alpha/b$. As expected, the Borel distribution (\ref{eq:Borel}) is recovered for $\alpha=0$. The distribution for $0<\alpha \leq b$ is different and can be summarized in the following observations: 
(\textit{i}) For all values $0\leq \alpha/b < 1$ the original power-law behavior in (\ref{eq:Borel}) is preserved below a chacateristic $N_T$ value $N_c \approx (1-N_b)^{1/(1-\alpha/b)}$ \cite{Saichev2005} but with a general exponent value  $\gamma_N^{\mathrm{low}}$.
(\textit{ii}) This effective power-law exponent $\gamma_N^{\mathrm{low}}$ observed below $N_c$ increases from the expected $1.5$ for $\alpha/b<0.5$ towards higher values $\gamma_N^{\mathrm{low}} \approx 2.0$ for $\alpha/b \approx 1$ (see Fig.~\ref{fig:pNt}.a).
The thick gray line in Fig.~\ref{fig:pNt}.b represents the Borel distribution for $N_b = 0.99$, indistinguishable to a power-law with $\gamma_N^{\mathrm{low}}$ within the range of observation. Since the characteristic scale is high ($N_c > 10^4$ for all $\alpha/b$), the exponents are used as a proxy for  $\gamma_N^{\mathrm{low}}$. The present results for $\gamma_N^{\mathrm{low}}$ agree with~\citeA{Saichev2005}, which predicted a transition from $\gamma_N^{\mathrm{low}} = 1.5$ for $b/\alpha<0.5$ to $\gamma_N^{\mathrm{low}}= 1+b/\alpha$ for $0.5\leq \alpha/b \leq 1$ (dashed green line in Fig.~\ref{fig:pNt}.a).
(\textit{iii}) The exponential regime for large trees found in the Borel distribution becomes fat-tailed with a power-law exponent $\gamma_N^{\mathrm{high}} = 1+b/\alpha$ for all the regime $0<\alpha < b$, reminiscent of the large scale regime in Eqs.~(\ref{eq:distroNb},\ref{eq:distroN}). This power-law regime can be observed in the distributions for $N_b= 0.30$ represented in Fig.~\ref{fig:pNt}.c.
The exponents  for $N_b = 0.30$ are used as a proxy for $\gamma_N^{\mathrm{high}}$ since the selected estimation interval is considerably above $N_c$ except for $\alpha=b$. 
(\textit{iv}) For $\alpha=b$, both exponents coincide and $N_c$ diverges. The power-law behavior has a single exponent ($\gamma_N \approx 2.0$) with an infinite domain.
This singular scale-free sampling is usually observed in field catalogs and incited the development of the more restrictive Branching Aftershock Sequence (BASS) model introduced by~\citeA{Turcotte2007}. In this singular solution, the value of $N_b$ only affects the proportion of singlets $\mathbb{P} (N_T = 0)$ (\ref{eq:distroN}), offsetting the distribution for all $N_T > 0$. This can be shown by comparing the faintest lines in Fig.~\ref{fig:pNt}.b and Fig.~\ref{fig:pNt}.c.
\\
\medskip
Other common topological observables can be inferred from the sampling distribution. Eq.~(\ref{eq:distroN}) establishes a relation between the ETAS parameters and the number of \textit{singlets}, by imposing $k=0$. Given that the branching process is GW, the same value provides the average fraction of leaves in trees and a good approximation to the average \textit{family size} $ \langle B \rangle$. The family size ($B$) is defined by~\citeA{Zaliapin2013b} as `\textit{the average number of offspring over all earthquakes in the family that have at least one offspring}'. This definition is mathematically identical to  $B :=\frac{N_T-1}{N_T-n_l}$, being $N_T$ the cluster size and $n_l$ the number of leaves. Notice that, for large clusters, $B\approx \left({1-\frac{n_l}{N_T}}\right)^{-1}$. Thus, the expected value of $B$ is directly related to the probability that an event is a leaf ($n_l/N_T$), which is equivalent to $P(N_i=0)$ or $g(0)$ in~\citeA{Zaliapin2013b}. 
Notice that $P(N_i=0)$ and $B$  depend on $N_b$ and the ratio $\alpha/b$ but are essentially independent of the number of events ($N_T$) in the standard ETAS model when $N_T$ is large enough.\\

\section{Generational depth and sizes}

\medskip
Some of the most commonly used topological properties of a GW process related to the generational depth of extinguished trees can be derived from the dualities between branching processes and random walks~\cite{Bennies2000,Pitman2006}. 
The concept of \textit{Harris path} \cite{Harris1951} is particularly useful for the topological analysis of aftershock sequences since it establishes a link between topological concepts such as size and depth with properties of a one-dimensional stochastic process.
The Harris path can be interpreted as a sorted exploration of the branching tree in the following way: Consider an event $i$ in the tree (parameterized as $t-1$ in the Harris path) of depth $d(t-1):=d_{i}$ that can either be a root ($d=0$)  or a triggered event from a parent with index $p$. We ask if this event $i$ has a new child $c$. If it does, we repeat the process with that first child event, now with $d(t) :=  d_c = d(t-1) +1$; if it does not, we decrease the generation by one ($d(t) :=  d_p = d(t-1) -1$), move back to the parent and ask if $p$ has another child unexplored by the Harris path. Starting from the mainshock or root of the tree, $d(t=1)=0$, all events $\lbrace i \rbrace$ are explored $\lbrace N_i + 1 \rbrace$ times before reaching $d(T)=-1$ at exactly $t=T\equiv 2N_T$.
Leaves are identified as local maximums in the profile. We find a leaf $l$ at $t$ if $ d({t})= d({t-1}) +1$ and $ d(t+1)= d({t}) -1$, and $l$ is explored exactly once by the Harris path. \\
\medskip
Let's consider, for a moment, that the tree is generated through a GW process  with a geometric offspring distribution. This is
$p_{1,N_b}(k)={N_b}^{k}(N_b+1)^{-k-1}$
 in terms of $N_b$. The probability of finding a new child, adding a step $d\to d+1$ in the Harris path, is independent of the number of previous children of the same parent and, therefore,  
the Harris path is equivalent to a random walk~\cite{Harris1951}. Hence, the depth profile is a diffusion process that scales with time as $t^{1/2}$. All magnitudes associated with charactersitic depths such as the maximum depth $D$ and the average leaf-depth ($\langle d_l \rangle$) are expected to scale with the size of the tree as $N_T^{1/2}$. Whilst the random walk analogy is only strictly valid in the geometric case, the Harris path of a GW with another offspring distributions is asymptotically equivalent to a random walk for large tree sizes as long as $N_b\lesssim 1$  and the variance is well defined ($0<\sigma^2<\infty$) \cite{Aldous1991b,Pitman2006}. This approximation is valid for the P-GW process, but not for the ETAS model when $\alpha/b>0.5$, rendering exponents values lower than 3 in Eq.~\ref{eq:distroN} and, hence, infinite variance.\\
\medskip

\begin{figure}
\includegraphics[scale=1]{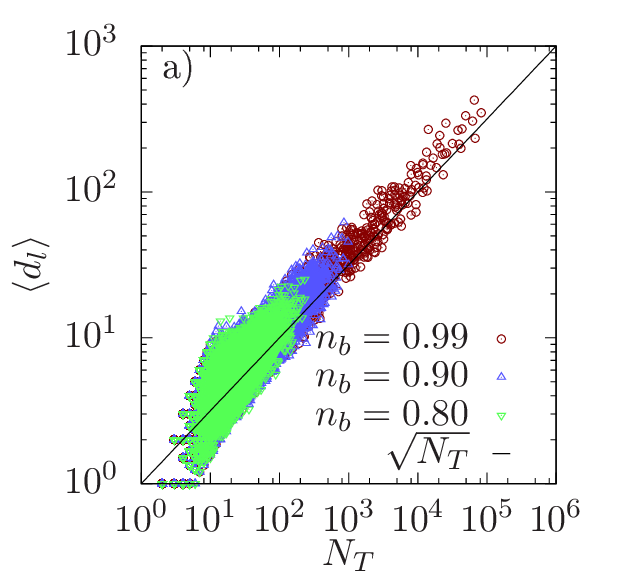}
\caption{\label{fig:GW}
a) Relation $d_l(N_T|n_b)$ of $N=10^6$ simulated P-GW trees for $n_b=0.8,0.9,0.99$. Black line represents the expected relation $\langle d_l | N_T\rangle \sim N_T^{0.5}$.
(don't need b)
}
\end{figure}

Fig.~\ref{fig:GW}.a shows the numerical validation of the diffusive assumption in the case of the P-GW. On average, the dependence between $d_l$ and $N_T$ follows the relation $\langle \langle d_l \rangle| N_T\rangle \sim N_T^{0.5}$, which extends to significantly low values of $N_b$. Although not shown here, a proportional relationship is also found in other characteristic depths such as the average depth of all events $\langle \langle d \rangle| N_T\rangle \sim N_T^{0.5}$ and average maximum depth $\langle \langle D \rangle| N_T\rangle \sim N_T^{0.5}$. Notice that, similarly as how a stochastic process cannot diffuse faster than a ballistic trajectory, the maximum depth cannot be larger than the tree size, forcing the limit $\langle d_l \rangle < N_T$ which biases the average values for small trees (usually $N_T<20$ as seen in Fig.~\ref{fig:GW}).\\

\medskip
\begin{figure}
\includegraphics[width=\textwidth]{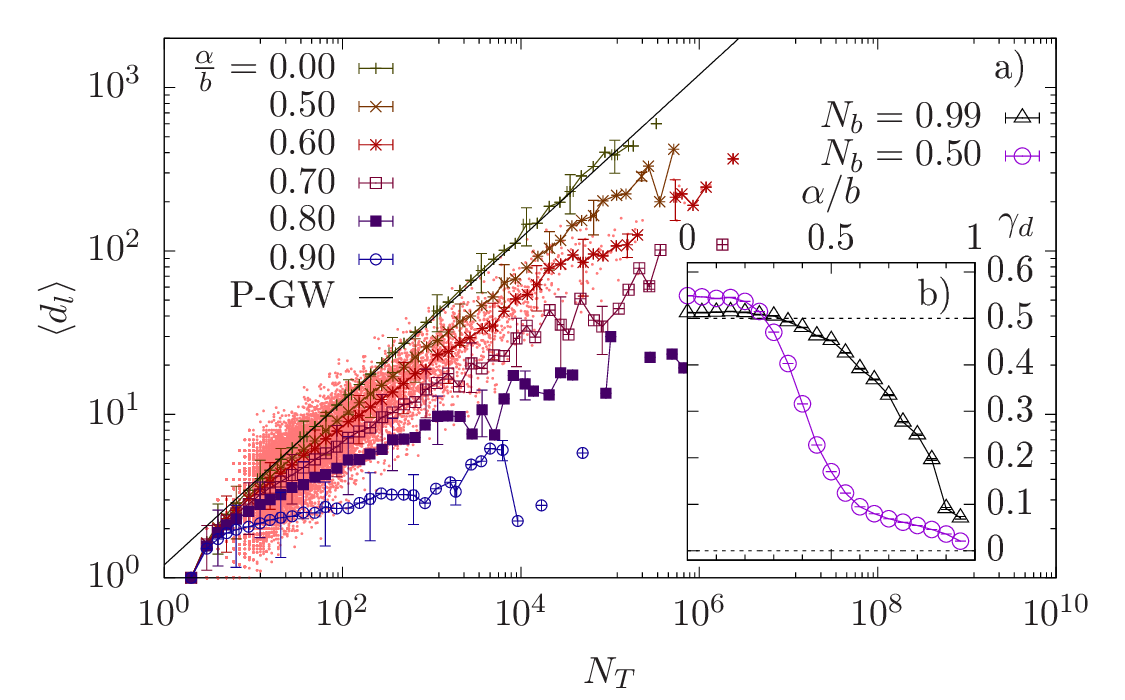}
\caption{\label{fig:dNt} a) Relation between average leaf-depths ($\langle d_l \rangle$) and tree size ($N_T$) for $N_b=0.99$. The scatter plot corresponds to trees sampled with $\alpha/b=0.60$. Lines show the conditional averages $\langle \langle d_l \rangle| N_T \rangle$ in independent windows of $N_T$. Error bars represent $\sigma$ of the conditional distribution. b) estimated exponent $\hat{\gamma}_d$ of the power-law relation (\ref{eq:dN}) within $30<N_T<1000$ for $N_b=0.99$ (triangles) and $N=0.50$ (circles) and different ratios $\alpha/b$.}
\end{figure}
The topological relations between depths and sizes are more complex in the ETAS model due to the power-law sampling (\ref{eq:distroN}). Here we only introduce the numerical results and leave the mathematical derivation, if possible, as an open question for future works.
Fig.~\ref{fig:dNt}.a shows the conditional average $\langle \langle d_l \rangle | N_T \rangle$ for different parameters $\alpha/b$. An scatter plot of the  data pairs ($\langle d_l \rangle$, $N_T$) for $\alpha/b=0.6$ and $N_b=0.995$ is also shown as example. 
The distribution of small trees exhibits the aforementioned limit $\langle d_l \rangle < N_T$, biasing the bivariate distribution for $N_T<20$. For larger trees,  an average power-law relation is observed beyond statistical fluctuations: 
\begin{equation}
\langle \langle d_l \rangle | N_T \rangle \propto N_T^{\gamma_d}.
\label{eq:dN}
\end{equation}
This power-law relation gets distorted for low values of $N_b$, although one is still able to estimate a reliable power-law relation for a shorter range.
The inset in Fig.~\ref{fig:dNt}.b shows the exponent values $\gamma_d$  fitted for trees with $30<N_T<1000$. The exponent appears to be consistent with the typical GW process ($\gamma_d=0.5$) for the relatively broad range $0<\alpha/b < 0.3$ but drops towards lower exponent values for larger ratios. 
This first regime for low $\alpha/b$ values is consistent with the asymptotic limit since the effective power-law exponent $\gamma_1$ in Eq.~\ref{eq:distroN} is higher than 3. For $0.5 <\alpha/b <1$, the drift from the diffusive relationship is apparent for all values of $N_b$. 
Close to $\alpha/b = 1$, the depth of triggering trees appear to be independent of $N_T$. 
The branching ratio $N_b$ changes the range of the distribution in $N_T$ as well as the dependence on the bivariate distribution. For low $N_b$ values, large events tend to have shallower trees than predicted by the diffusive relationship. As a consequence, the effective exponent $\gamma_d$ depends on $N_b$ for intermediate values of $\alpha/b$ but almost coincide for the extreme cases $\alpha < 0.3b$ and $\alpha \approx 1$ . The exponent transition is steeper for low $N_b$ values. 
\section{Discussion}
The analytical and numerical results obtained here from the branching process models highlight three direct measurements that stand out to characterize the topology of epidemic aftershock processes: the direct triggering distribution $p_1(k)$, the relation $\langle d_l \rangle$, $N_T$ and the size distribution $p_N(K)$. Their joint analysis can verify whether the relation between $ \langle d_l \rangle$ and $N_T$ corresponds to the expected behavior of the specific branching model hypothesis and an ETAS model with independently fitted parameters $k_0$, $b$ and $\alpha$. Scalar measurements such as the fraction of singlets or leaves, the family size $B$, and the bulk average leaf-depth $\langle \langle d_l \rangle \rangle $, which entangle both dependencies in $N_b$ and $\alpha/b$, can mislead the validation or fitting of a branching model. However, the expected relationships between such scalar measurements, such as the fraction of leaves and singlets, can serve for preliminary rejection tests.\\
\medskip
The ETAS model can potentially explain the topological variability of triggering trees reconstructed from seismological catalogs \cite{Zaliapin2013b} and acoustic emission experiments \cite{Davidsen2017}. 
Provided the validity of the branching model approach, unequivocal relations exist between the topological properties and the parameters of aftershock production, determined by $N_b$ and $\alpha/b$ in the ETAS model (\ref{eq:distroN}). 
Assuming independence between the values of $n_b$ ---guaranteed in the present study because of the independence of $m$--- all topological information is contained in the distribution of $n_b$ (as Eq.~\ref{eq:distroNb} in the  ETAS). 
In general, the same relations can be extrapolated to other GW models with well defined $p_1(k)$. 
Beyond strict GW processes, the same results would be locally valid in models implementing spatio-temporal variations of the ETAS parameters such as the spatially-variating ETAS model (SVETAS) introduced by~\citeA{Nadan2017}. In that case, the topological properties would depend on the local distribution of $N_b$ and $\alpha/b$. A global evaluation would report a spread in the ($\langle d_l \rangle$,$N_T$) space with a blurred power-law relationship, as reported by~\citeA{Zaliapin2013b} in southern California.
 \\
\medskip
The ETAS model is a peculiar case of GW processes where the aftershocks sampling is power-law~\cite{Saichev2005}, leading to a natural cluster classification in the terms presented by~\citeA{Zaliapin2013b}. The results shown in Fig.~\ref{fig:dNt} validate the hypothesis that swarms, like bursts, can appear as a consequence of event-event triggering processes, i.e. aftershocks, represented as one-to-one causal links in branching processes~\cite{Zaliapin2013b}. 
The topological properties of the trees used for the classification of swarms and bursts ---and, in particular, the exponent $\gamma_d$--- differ depending on the parameter ratio $\alpha/b$ and the branching ratio $N_b$.  
Such classification is noticeably sharp in the parameter space for low $N_b$ ( Fig.~\ref{fig:dNt}.b) but smooth for high branching ratios ($ N_b \lesssim 1$). The two classes are found in the extreme cases $\alpha = b$ and $\alpha=0$.
When $\alpha < 0.5 b$, the P-GW limit is recovered and trees grow as swarms, forming relatively slender tree structures, with $\gamma_d \approx 0.5$. 
In the opposite case scenario, when $\alpha \approx b$, only strong events are likely to generate aftershocks. Because the branching ratio is fixed, most of the triggered activity for $\alpha \approx b$ is due to the few stronger events, which, as consequence, are more likely to be background events. 
The first generation offspring of this strong event is unlikely to generate aftershocks of their own, rendering spray-like short tree sequences and star-shaped spatial structures characteristic of burst-like activity  \cite{Zaliapin2013b}. 
Although the transition is not sharp in the parameter space, the empirical $\alpha/b$ values are typically close to one in seismological catalogs. Hence the separation between $\alpha/b \approx 1$ and $\alpha/b < 1$ is a natural choice to define the classification between burst-like and swarm-like clusters.
The GW model does not expect values $\gamma_d> 0.5$. Significantly higher exponent values might indicate memory in the branching process, which cannot be modeled as a GW.
\\
\medskip
Overall, the ETAS model establishes a clear relationship between the topological properties and the ratio between the Gutenberg-Richter exponent $b$ and the productivity exponent $\alpha$. Swarms are resulting particularly problematic when validating the ETAS model through this relationship.
On the one hand, \citeA{Zaliapin2013b} had shown how the ETAS model parameterized with field estimations fails to predict the observed topology of swarms in southern California, where $\alpha \approx b$ and $\gamma_T \sim 1$.  The distribution of swarm-like activity in hot areas~\cite{Zaliapin2016} is consistent with the power-law behavior in Eq.(\ref{eq:distroN}), rendering the expected value for $b/\alpha = 1$. Hence, the swarm-like shape of the trees is inconsistent with the ETAS model given the fitted parameters (see Fig.~\ref{fig:dNt}).
On the other hand, cluster reconstruction techniques are based on the homogeneous Poisson null-hypothesis~\cite{Zaliapin2013a} and do not take into account spatio-temporal variations due to exogenous geological or anthropogenic processes, i.e. when $\mu_0(t,\mathbf{r})$ has a dependence in both space and time. This is the case, for example, of episodic volcanic \cite{Roberts2016}, natural geothermal \cite{Gaeta1998}, or human-induced \cite{Ellsworth2013} seismicity, or even tectonic seismicity in the presence of seasonal variations \cite{Ueda2019}. Spatio-temporal correlations in such settings are not necessarily a consequence of a history-dependence and might disrupt the performance of cluster detection techniques, which will overestimate triggering relationships. Precisely, the random linking of uncorrelated events would lead to the generation of a P-GW process, compatible with the tree shapes reported by \citeA{Zaliapin2013b}. In particular, geothermal systems  reproducing P-GW processes might not represent actual triggering, but rather exogenous variations in the background rate. Studies focused on well confined episodes of seismic activity are advised to validate the results.\\
\medskip
Discrepancies between model and data might be corrected by modifications of the ETAS models implementing more sophisticated field observations. 
Magnitude-magnitude correlations \cite{Lippiello2008,Davidsen2011} or depth-dependent $m$-distributions would, in general, fall outside the GW category. However, this is not a rule of thumb. In particular, recent observations suggest a simple distinction in the Gutenberg-Richter exponent for aftershocks ($b_{AS}$) and mainshocks ($b$) \cite{Gu2013,Davidsen2016,Davidsen2017}. This specific modification falls still within the GW category. The relation between topological properties would, in that case, depend on $k_c$, $\alpha$, $b_{AS}$ and $b$.
\\
\medskip
\medskip
%
Finally, any empirical study on the topological properties should account for the fact that natural catalogs have a limited spatial and temporal range. Even if the branching approach is valid, the concept of leaf and root are ill-defined in spatially confined or finite time series~\cite{Zhuang2002}. Events misclassified as mainshocks might be actually triggered by remote or ancestral events~\cite{Elst2017} and apparent leaves might generate aftershocks outside the observational range. The results can be especially biased for the analysis of short catalogs, considering the power-law kernels in $ \Psi_t$  and $\Psi_r$ of the ETAS model.\\
\medskip

\section{Conclusions}

Here, we have considered the branching model as a valid representation of aftershock sequences and other triggering processes. The reconstruction of triggering trees accounting for all event-event correlations opens new perspectives to learn about the seismogenic mechanisms behind aftershocks and improve our current forecasting techniques. 
We have revisited, and added to, the expected topological properties for the standard Epidemic-Type Aftershock Sequence model (ETAS), which is interpreted as a fat-tailed Galton-Watson process, extended from the Poisson Galton-Watson process (P-GW) which is recovered as a particular case. This list of properties had been proposed in previous works to be helpful to characterize and classify aftershock sequences. Specifically, this analysis serves to distinguish between swarms and bursts \cite{Zaliapin2013b,Zaliapin2016}.\\
\medskip
All the singular properties of the ETAS model within the GW category derive from the power-law distribution of branching ratios $n_b$ leading to a similar power-law in $p_1(k)$. In particular, all topological properties depend only on two parameters: the average branching ratio $N_b$ and the ratio between exponents $\alpha/b$. Since the P-GW is recovered as a limiting case of the ETAS model we observe a transition in the distribution of tree sizes ($N_T$). The $N_b$ controls the characteristic $N_T$ marking the transition between two distinct power-law exponents  ($\gamma_N^{\mathrm{low}}$ and $\gamma_N^{\mathrm{high}}$) , which coincide  to $\gamma_N^{\mathrm{low}}=\gamma_N^{\mathrm{high}}=2$ for $\alpha=b$, and recovers the Borel distribution for $\alpha=0$. Characteristic generational depths are strongly dependent on $\alpha/b$ once the power-law tail has a significant statistical weight ($\gamma_1<3$). The average leaf-depth of a tree ($\langle d_l \rangle$) has a power-law dependence with $N_T$ for high branching ratios or low exponent ratios $\alpha/b$. The exponent of this power-law relation coincides with $\gamma_d=0.5$, typical in well defined GW processes, for $\alpha=0$ and decreases for values $ b \lesssim 2 \alpha$,  vanishing to zero at the limit $\alpha = b$. For low branching ratios this transition in the exponent gets sharper and occurs at lower values, for example $ b / \alpha \sim 0.3$ for $N_b=0.5$. This numerical result interprets the separation between bursts and swarms as a phenomenological observation based upon the common $\alpha \approx b$ found in nature. A regional analysis of the ratio is required to validate the exponent dependence on the $\alpha/b$ ratio. In any case, the results in the ETAS model prove that the topological structure of swarms, as bursts, can be explained  as an event-event triggering processes, i.e. aftershocks, represented by one-to-one causal links.\\
\medskip
In general, this study can be used to validate the ETAS model for the description of the triggering processes associated with tectonic and induced seismicity as well as acoustic emission experiments \cite{Benioff1951,Hirata1987,Baro2013,Ribeiro2015,Costa2016,Davidsen2017} and micromechanical models \cite{Yamashita1987, Dieterich1994, Hainzl1999, Lyakhovsky2005, Jagla2010, Zhang2016, Baro2018a}. A rejection of the ETAS model from the topological properties of the triggering trees might indicate more sophisticated epidemic processes, involving magnitude-magnitude correlations or depth dependencies. On the contrary, the validation of the  ETAS model would set a step forward in the testing and development of new micromechanical models implementing seismogenic mechanisms of aftershocks and other event-event triggering mechanisms.

%
%
%
%
%
%
%
%

\acknowledgments
Thanks to A. Corral, K. Wiese and I. Zaliapin for fruitful discussions.
Thanks to AXA Research Fund for financial support through the project RheMechFail.


%
%

\bibliography{biblioETASGW}{}

%
%
%
%
%

\end{document}